\documentclass[aps,preprint,superscriptaddress,groupedaddress]{revtex4-1}
\usepackage{graphicx}
\usepackage{epsfig}
\usepackage{amssymb}
\usepackage{mathrsfs}
\usepackage{amsmath}
\usepackage{color}
\usepackage{latexsym}
\usepackage{amsfonts}
\usepackage{hyperref}
\usepackage{subfigure}
\usepackage{wasysym}
\usepackage{dcolumn}
\usepackage{bm}

\begin{document}

\selectfont

\title{Critical behavior of a water monolayer under hydrophobic confinement}

\author{Valentino Bianco}
\affiliation{Departament de F\'isica Fonamental, 
Universitat de Barcelona, Martí i Franquès 1, ES-08028 Barcelona, Spain}
\email{gfranzese@ub.edu}

\author{Giancarlo~Franzese}
\affiliation{Departament de F\'isica Fonamental, 
Universitat de Barcelona, Martí i Franquès 1, ES-08028 Barcelona, Spain}

\date{17 March 2014}

\definecolor{corr14okt}{rgb}{0,0,1} 
\definecolor{moved}{rgb}{0,0,0}

\maketitle

\noindent{\bf The properties of water can have a strong dependence on the confinement. 
Here, we consider a water monolayer nanoconfined between hydrophobic parallel walls under
conditions that prevent its crystallization. We investigate, by
simulations of a many-body coarse-grained water model, how the properties of the liquid are
affected by the confinement. We show, by studying the response functions and the correlation length  and by performing finite-size scaling of the appropriate order parameter,  that at low temperature
the monolayer undergoes a liquid-liquid
phase transition ending in a 
  critical point in the universality class of the
  two-dimensional (2D) Ising model. 
Surprisingly, by reducing the linear size $L$ of the walls,
  keeping the walls separation $h$ constant, we find a 2D-3D crossover for
  the universality class of the liquid-liquid critical point 
for $L/h\simeq 50$, i.e. for a
  monolayer thickness that is small compared to its
  extension. This result is drastically different from what is
  reported for simple liquids, where the crossover occurs for
  $L/h\simeq 5$, and is consistent with experimental results and
  atomistic simulations. We shed light on these findings 
showing that they are a consequence of  the strong cooperativity
and the low coordination number of the hydrogen bond network that 
characterizes water.} 
\\\\\\PACS number 64.70.Ja, 65.20.-w, 68.15.+e
  
The study of nanoconfined water is of great interest for applications
in nanotechnology and nanoscience \cite{majumder}. The confinement of water in quasi-one or two
dimensions (2D) is leading to the discovery of new and controversial
phenomena in experiments \cite{majumder,
  Whitby2007,Zhang2011, soper, han2010} and simulations \cite{Whitby2007, faraudo-bresme, shr}. 
Nanoconfinement, both in hydrophilic and hydrophobic materials, can keep water in the liquid phase at  temperatures as low as 130 K at ambient pressure \cite{Zhang2011}.
At these temperatures  $T$ and pressures $P$ experiments cannot probe
liquid water in the bulk, because water freezes faster then the
minimum observation time of usual techniques,  resulting in an
experimental ``no man's land'' \cite{Mishima1998}. Nevertheless, new
kind of experiments \cite{nillson1,taschin} and numerical simulations  \cite{llcp} can access this region,
revealing interesting phenomena in the metastable state.
In particular, Poole et al.  found, by molecular dynamics simulations
of supercooled water, a liquid-liquid critical point (LLCP), in the
``no man’s land'', at the end of a first--order liquid-liquid phase
transition (LLPT) line between two metastable liquids phases with
different density $\rho$: the high-density liquid (HDL) at higher $T$
and $P$, and the low-density liquid (LDL) at lower $T$ and $P$
\cite{llcp}.  The presence of a LLPT is experimentally observed in
other systems \cite{anomalous-exp, exp1, exp2, exp3, exp4, exp5, exp6,
  exp7,exp8, mcmillan}, consistent with theoretical models fitted to water
experimental data \cite{Anisimov, holten1,nillson2}, and is recovered by
simulations of a number of models of water \cite{llcp,LLCPnew, llcp1,
llcp2, llcp3, llcp4_bis, llcp5,FMS2003} and other anomalous liquids
\cite{anomalous-sim, an1, an2, an3, an7,
  an8}. Alternative ideas, and their differences, have been
discussed \cite{sf,Stokely2010, Limmer,llcp4,palmer}, and it has been debated if
experiments on
confined water in the ``no man's land'' can  be a way to
test these ideas \cite{Zhang2011}, motivating several
theoretical works  \cite{gallo-rovere}. 

\begin{figure}
 \includegraphics[scale=0.25]{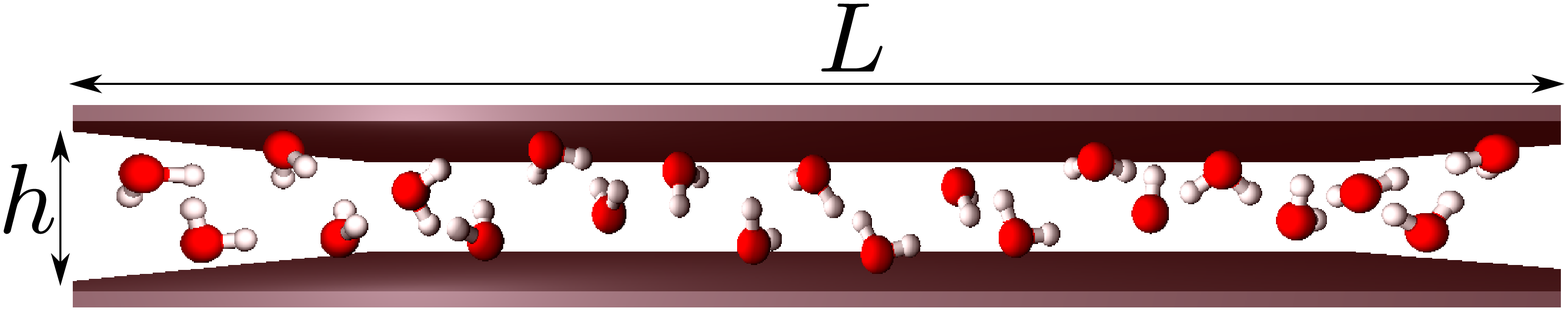}
\caption{Schematic view of a section of the water monolayer
confined between hydrophobic walls of size $L\times L$ separated by $h\approx 0.5$~nm.}
\label{schematic}
\end{figure}

Here, to analyze the thermodynamic properties of water in confinement
we consider a water monolayer between
hydrophobic walls of area $L^2$ separated by $h\approx 0.5$~nm
(Fig.~\ref{schematic}).  Atomistic simulations \cite{shr} show that  water under these conditions does not
crystallize, but arranges in a disordered liquid layer, whose
projection on one of the surfaces has square symmetry, with each water
molecule having four nearest neighbors (n.n.). The molecules maximize
their intermolecular distance by adjusting at different heigths with respect to
the two walls. 

We adopt a {\it many-body} model that reproduces water properties
\cite{FMS2003,Stokely2010,KFS2008,dls,MSSSF2009,bb,bbb, Mazza, fbfood2013}. We simulate $\sim 10^5$ state points, each with
statistics of $5\times 10^6$ independent calculations, for systems with
$N=2.5\times10^3, \dots, 1.6\times10^5$ water molecules at constant $N$, $P$
and $T$, using a cluster Monte Carlo algorithm \cite{MSSSF2009,bb,bbb}, 
for a wide range of $T$ and $P$. All quantities are calculated in internal units, as described
in the Methods section.

\medskip
\bigskip
\noindent{\bf Results}

\begin{figure}

\includegraphics[scale=1,bb=70 300 612 720,clip=true]{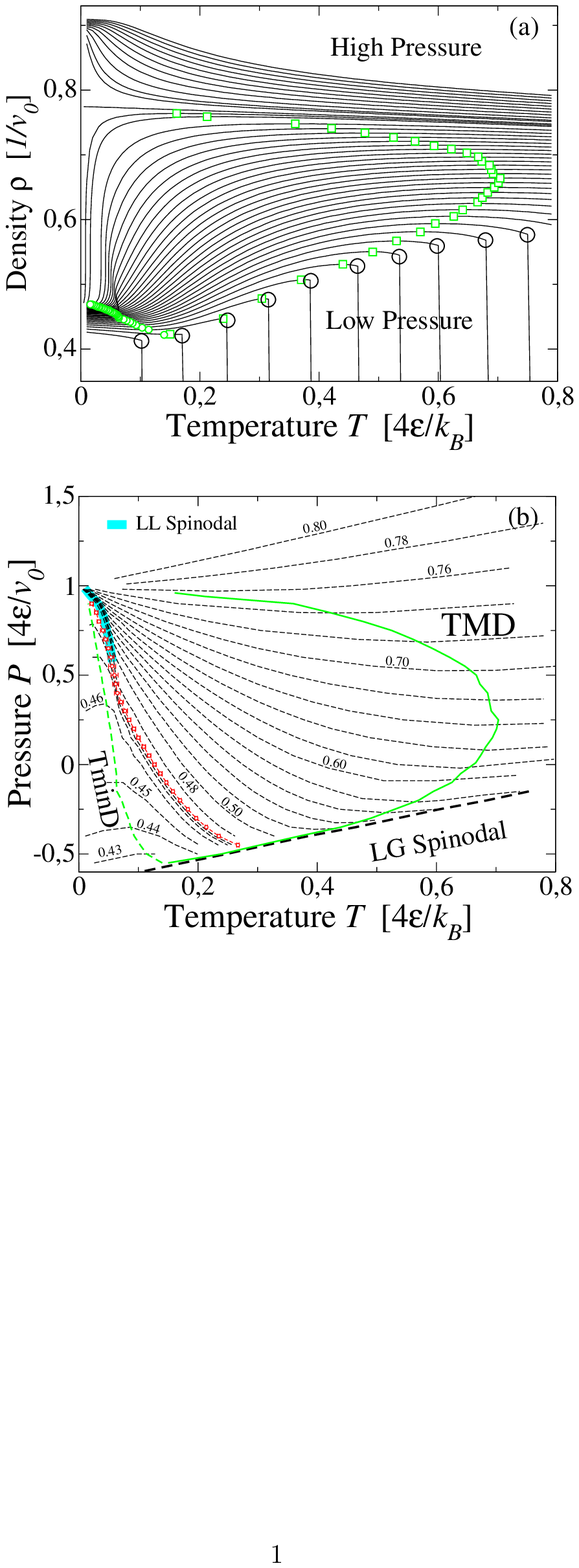}
\caption{(a) Isobaric density variation for $10^4$ water
  molecule. Lines join simulated state points ($\sim 150$ for each
  isobar). $P$ increases from $-0.5$ (bottom curve) to $1.5$ $(4\epsilon)/v_0$  (top
  curve). Along each isobar we locate the maximum $\rho$ (green squares
at high $T$) and the minimum $\rho$ (green small
  circles at low $T$) and the liquid-gas spinodal (open large circles
  at low $P$).  
(b) Loci of TMD, TminD, liquid-gas spinodal and liquid-liquid spinodal
in $T-P$ plane. Dashed lines with labels represent the isochores of the system from
$\rho v_0=0.43$ (bottom) to $\rho v_0=0.80$ (top). Dashed lines without labels represent intermediate isochores. TMD and TminD correspond to
the loci of minima  and maxima, respectively, along isochores in the
$T-P$ plane. We estimate the critical isochore at $\rho v_0\sim0.47$
(red circles). All the isochores with $0.47<\rho v_0<0.76 $  intersect with
the critical isochore for $Pv_0/(4\epsilon)\geq 0.5$  along the LL spinodal (tick
turquoise) line. 
} 
\label{density}
\end{figure}

We calculate the density $\rho\equiv N/V$ of the system as function of
$T$ along isobars. For a broad range of $P$, we find a maximum and a
minimum of density along each isobar (Fig. \ref{density}a) according
to experimental evidences for bulk and confined water
\cite{Mallamace-density-min}. These maxima and minima identify, for
each $P$, the temperature of maximum density (TMD) and the temperature of
minimum density (TminD). The TMD locus merges the TminD line as in experiments \cite{Mallamace-density-min} and other models \cite{Poole2005}.

At low $T$ a discontinuous change
in $\rho$ is observed for $1>P v_0/(4\epsilon)\geq 0.5$, where the parameters $v_0$ and $\epsilon$ are explained in the Methods section, as it would be expected in
correspondence of the HDL-LDL phase transition. At very high pressures
($Pv_0/(4\epsilon) > 1$) the system behaves as a normal liquid, with monotonically
increase of $\rho$ upon decrease of $T$. 

We estimate the liquid-to-gas (LG) spinodal at $Pv_0/(4\epsilon) < 0$ for low $T$
(Fig. \ref{density}) as the temperature along an isobar at which we
find a discontinuous jump of $\rho$ to zero value by heating the system. The LG spinodal
identifies the locus of the stability limit of liquid phase with
respect to the gas phase: at pressures below the LG spinodal in the $P-T$ plane is no longer possible to
equilibrate the system in the liquid phase. The LG spinodal continues
at positive pressures ending in the LG critical point (data not
shown). We observe that the TMD line approaches the LG spinodal, without touching it (Fig. \ref{density}).
We recover the LG spinodal also as envelope of isochores
(Fig. \ref{density}b).

We find a second envelope of isochores at lower $T$ and higher $P$,
pointing out the liquid-to-liquid (LL) spinodal. Indeed, the two
spinodals associated to the LLPT, i.e. the HDL-to-LDL spinodal and the
LDL-to-HDL spinodal, collapse one on top of the other and are
indistinguishable within our numerical resolution. Nevertheless, we
clearly see that isochores are gathering around the points ($T k_B/(4\epsilon)\sim
0.06$, $P v_0/(4\epsilon)\sim0.5$ ) and ($T k_B/(4\epsilon)=0$, $P v_0/(4\epsilon)=1$), where $k_B$ is the Boltzmann constant, marking two possible critical regions (Fig. \ref{density}b).

\begin{figure}
\includegraphics[scale=1,bb=70 360 612 720,clip=true]{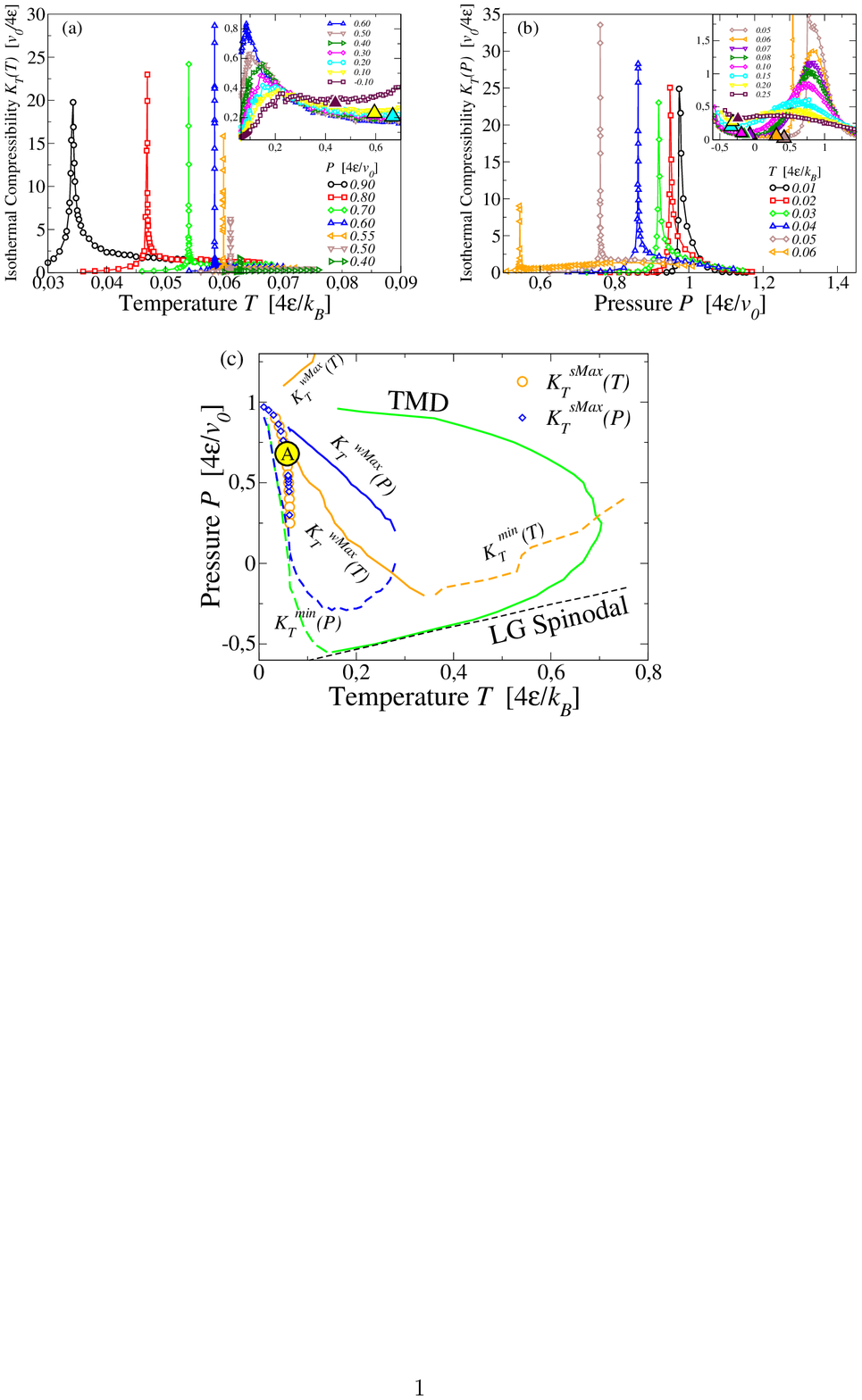} 
\caption{
(a) Loci of strong maxima ($K_T^{\rm sMax}(T)$), weak maxima ($K_T^{\rm wMax}(T)$
in the inset) and  minima ($K_T^{\rm min}(T)$ marked with large triangles in the inset) 
along isobars for $K_T(T)$.
(b) Loci of strong maxima ($K_T^{\rm sMax}(P)$), weak
maxima ($K_T^{\rm wMax}(P)$ in the inset) and minima ($K_T^{\rm
  min}(P)$ marked with large triangles in the inset)
along isotherms. 
The weak maxima merge with minima.
(c) Projection of
extrema of $K_T$ in $T-P$ plane. The strong   maxima (symbols), weak
maxima (solid lines) and minima (dashed lines) of $K_T(T)$ (orange)
and $K_T(P)$ (blue) form loci in $T-P$ plane that relate to each other
and intersect with the TMD line following the thermodynamic relations
discussed in the text. 
The large yellow circle with label A identifies the region where
$K_T^{\rm sMax}(T)$ and $K_T^{\rm sMax}(P)$ converge and display the
largest maxima, consistent with the occurrence of a critical point in
a finite-size system. 
Symbols not explained here are as in Fig.~\ref{density}.
} 
\label{kt_loci}
\end{figure}

We calculate the isothermal compressibility by its definition
$K_T\equiv- (1/\langle V \rangle)\left(\partial \langle V \rangle
  /\partial P\right)_T$ and by the fluctuation-dissipation theorem
$K_T= \langle\Delta V^2\rangle /k_BTV$ along isobars,
$K_T(T)$,
and along isotherms,
$K_T(P)$
(Fig. \ref{kt_loci}), where $\langle V \rangle\equiv V$ is the average volume and
$\langle \Delta V^2 \rangle$ the volume fluctuations.
We find two loci of extrema for each quantity $K_T(T)$ and $K_T(P)$:
one of strong maxima and one of weak maxima. The loci of strong
maxima in $K_T(T)$ and $K_T(P)$, respectively $K_T^{\rm sMax}(T)$ and
$K_T^{\rm sMax}(P)$, overlap within the error bar with the LL
spinodal. The maxima $K_T^{\rm sMax}(T)$ and $K_T^{\rm sMax}(P)$
increase in the range of $Pv_0/(4\epsilon)\in [0.55;0.6]$ and $Tk_B/(4\epsilon)\in [0.05;0.06]$
(Fig. \ref{kt_loci}), consistent with the existence of a critical
region. The stronger maxima disappear 
for $Pv_0/(4\epsilon) < 0.4$.

We find also loci of weak maxima, $K_T^{\rm wMax}(T)$ and $K_T^{\rm
  wMax}(P)$ and minima $K_T^{\rm min}(T)$ and $K_T^{\rm min}(P)$. The
loci of weak extrema  and minima of $K_T(T)$ and $K_T(P)$ do not coincide in the $T-P$
plane. 
The locus of weak maxima along isotherms  $K_T^{\rm wMax}(P)$ merges
with the locus of minima $K_T^{\rm min}(P)$ at the point where the
slope of both loci is $\partial P/\partial
T\rightarrow\infty$. Furthermore, both loci approach to the LL
spinodal at high $P$. 
The locus of weak maxima along isobars $K_T^{\rm wMax}(T)$ approaches
the LL spinodal where $K_T$ exhibits the strongest maxima, and merges
with the locus of minima $K_T^{\rm min}(T)$ where the slope of both
loci is $\partial P/\partial T\rightarrow0$ (data at high $P$ and $T$
not shown in Fig \ref{kt_loci}). This locus intersects the TMD at its
turning point. Indeed, as reported in Ref. \cite{sf} and in the
Methods section, the temperature derivative of isobaric $K_T$ along
the TMD line is related to the slope of TMD line 
\begin{equation}\label{rel1}
\left(\dfrac{\partial K_T}{\partial T}\right)_{P,~\rm TMD}=
\dfrac{1}{V} \dfrac{(\partial^2 V/\partial T^2)_{\rm TMD}}{ (\partial
  P/\partial T)_{\rm TMD}} 
\end{equation} 
where all the quantities are calculated along
the TMD line.
Hence the locus of extrema in $K_T(T)$, where $(\partial K_T/\partial
T)_{P}= 0$,  crosses the TMD line where the slope $(\partial
P/\partial T)_{\rm TMD}$ is infinite.  
We observe also that the weak maxima of $K_T(T)$ and $K_T(P)$ increase
as they approach the 
LL spinodal. All loci of extrema in $K_T$ are summarized in
Fig. \ref{kt_loci}.  
                                             
\begin{figure}
\includegraphics[scale=1,bb=70 360 612 720,clip=true]{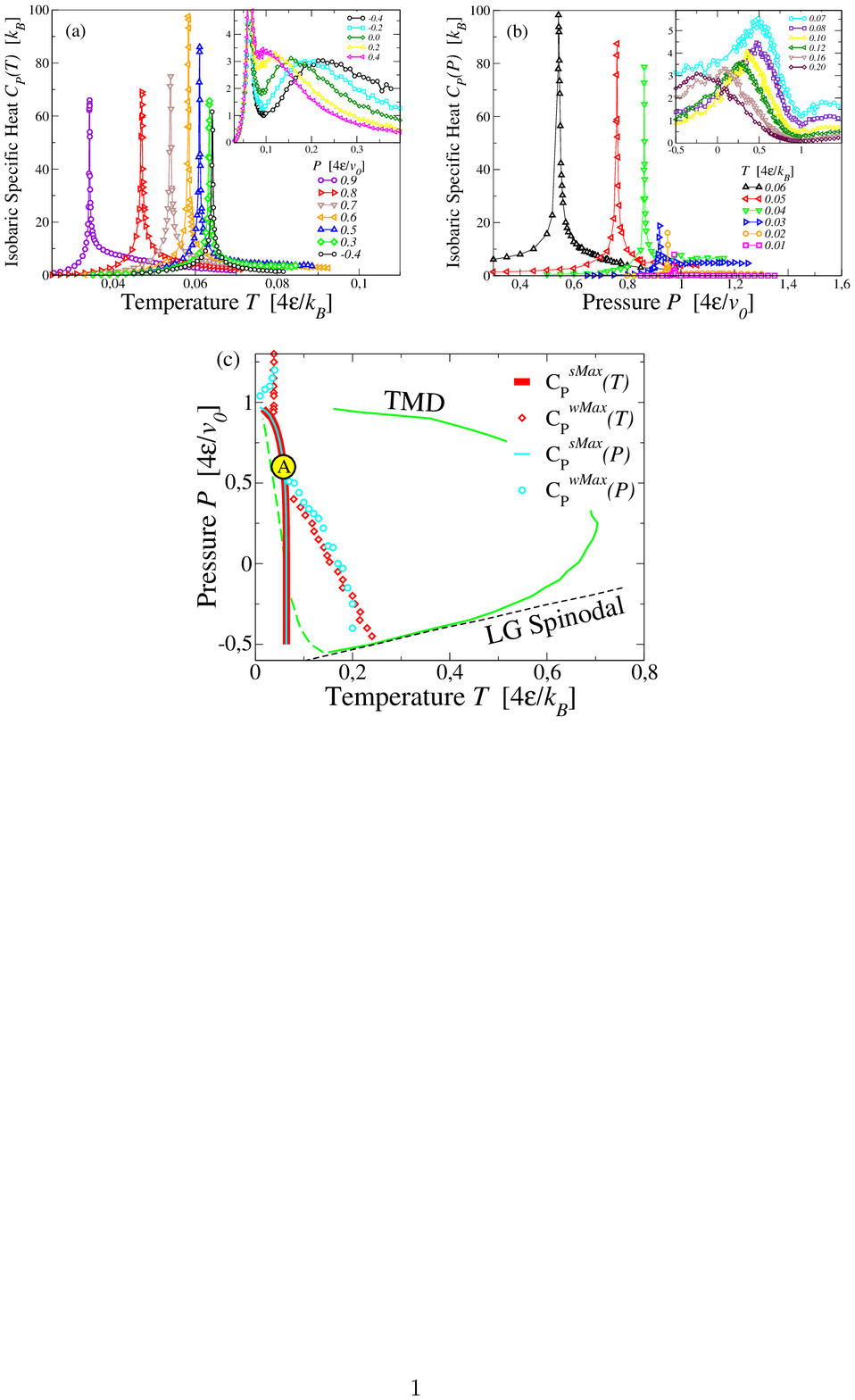}
\caption{
(a) Loci of strong maxima 
($C_P^{\rm sMax}(T)$) and weak
maxima ($C_P^{\rm wMax}(T)$ in the inset) along isobars for $C_P$.
(b) Loci of strong maxima ($C_P^{\rm sMax}(P)$) and weak
maxima ($C_P^{\rm wMax}(P)$ in the inset) along isotherms. 
(c)  Projection of $C_P$ maxima in $T-P$ plane.  
The large circle with A identifies the region where $C_P$ shows the
strongest maximum.
Symbols not explained here are as in Fig.~\ref{density}.}
\label{cp_loci}
\end{figure}

Next we calculate the isobaric specific heat $C_P\equiv \left(\partial
  \langle H \rangle /\partial T\right)_P =\langle\Delta
H^2\rangle/k_BT$ along isotherms and isobars, where $\langle H
\rangle\equiv \langle \mathscr{H} \rangle +P \langle V \rangle$ is the
average enthalpy, $\mathscr{H} $ is the Hamiltonian as defined in the
Methods section, 
$\langle \Delta H^2 \rangle$ is the enthalpy
fluctuations (Fig. \ref{cp_loci}).  
We find two maxima at low $P$ separated by a minimum. At high-$T$ the maxima are broader and
weaker than those  at low-$T$. As discussed in Ref. \cite{Mazza},  the maxima at high $T$ are related to maxima in
fluctuation of the HB number $N_{\rm HB}$, while the maxima at low $T$
are a consequence of maxima in fluctuations of the number $N_{\rm
  coop}$  of cooperative HBs. The lines of strong maxima at constant
$P$ and constant $T$, respectively  $C_P^{\rm sMax}(T)$ and $C_P^{\rm
  sMax}(P)$, overlap for all the considered pressures, and both maxima
are more pronounced in the range $Pv_0/(4\epsilon) \in [0.5,0.6]$ and $Tk_B/(4\epsilon)\in
[0.06,0.07]$. The weak maxima $C_P^{\rm wMax} (P)$ and $C_P^{\rm
  wMax}(T)$  increase approaching the LL spinodal and have their
larger maxima at the state point where they converg to the strong
maxima, consistent with the occurrence of a critical point for a
finite system (Fig. \ref{cp_loci}). The lines of weak maxima overlap
for all positive pressures, branching off at negative pressures. At
negative pressures, the locus  $C_P^{\rm wMax}(P)$ bends toward the
turning point of the TMD line, as discussed in Methods section and in
Ref. \cite{Poole2005}. Indeed, according to the relation  
\begin{equation}\label{rel2}
 \left(\frac{\partial C_P}{\partial P}\right)_{T, ~\rm TMD}=T\left(\frac{\partial P}{\partial T}\right)_{\rm TMD}\left(\frac{\partial^2V}{\partial P\partial T}\right)_{\rm TMD},
\end{equation} 
in case of intersection between the locus of extrema $(\partial C_P/\partial P)_T=0$ and the TMD line, it results that $(\partial P/\partial T)_{\rm TMD}=0$.  Note that, as we explain in the Methods section, the relation (\ref{rel2}) does not imply any change in the slope of the TminD line at the intersection with the locus of $(\partial C_P/\partial P)_T=0$.

\begin{figure}
\includegraphics[scale=1,bb=70 360 612 720,clip=true]{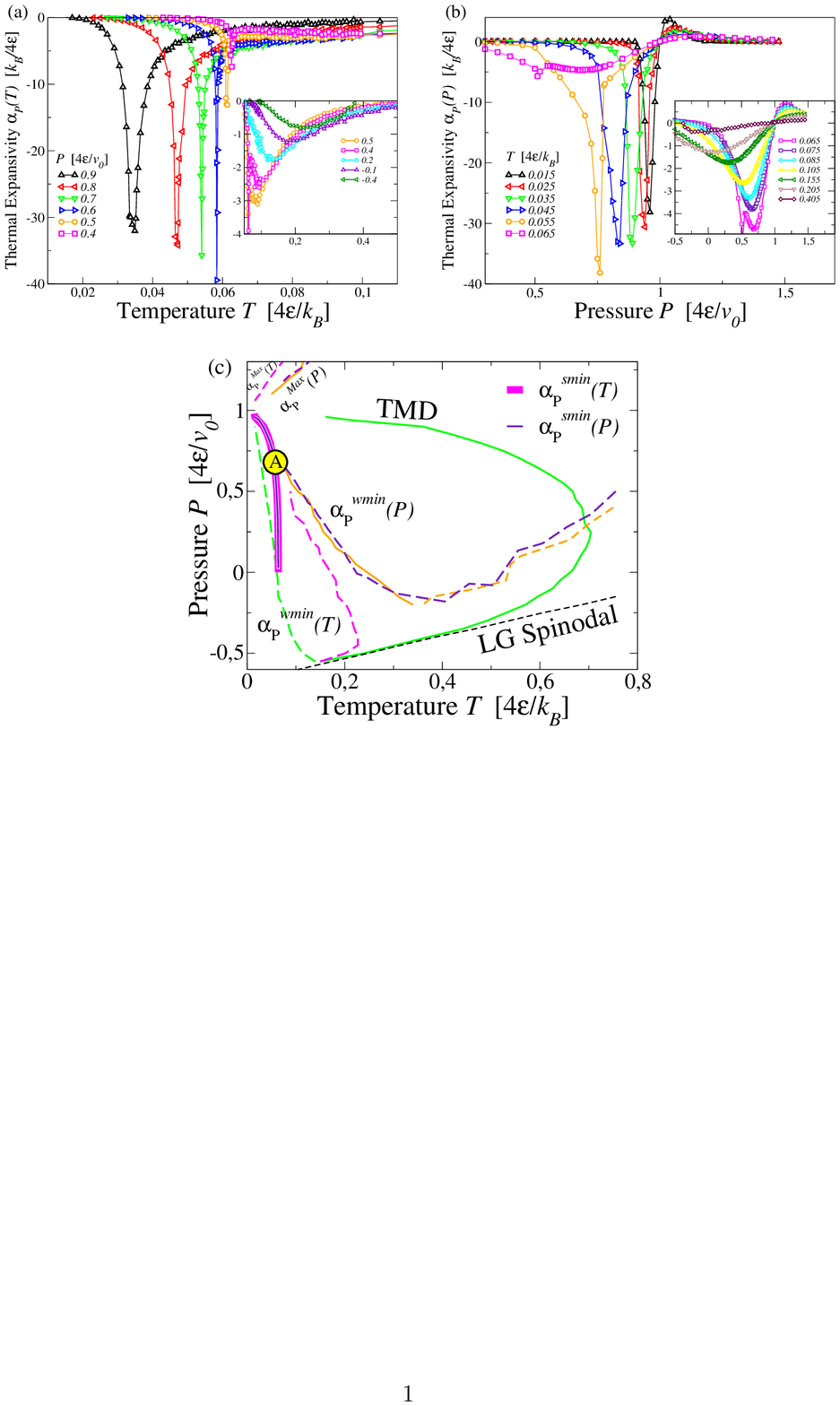}
\caption{ 
(a) Loci of strong minima of 
($\alpha_P^{\rm smin}(T)$) and
weak minima  ($\alpha_P^{\rm wmin}(T)$  in the inset) along isobars
for $\alpha_P$.
(b) Loci of strong  minima ($\alpha_P^{\rm smin}(P)$)
and weak extrema ($\alpha_P^{\rm Max}(P)$ and $\alpha_P^{\rm wmin}(P)$
in the inset) along isotherms. 
(c) Projection of 
$\alpha_P$ extrema  in $T-P$ plane. 
Orange lines are the loci of weaker extrema $K_T^{\rm wMax}$ and
$K_T^{\rm min}$. The large circle with A identifies the region where
the divergent minimum in $\alpha_P$ is observed. 
Symbols not explained here are as in Fig.~\ref{density}.
}
\label{alpha_loci}
\end{figure}

We calculate also the thermal expansivity $\alpha_P\equiv (1/\langle V
\rangle)\left(\partial \langle V \rangle /\partial T\right)_P$ along
isotherms and isobars (Fig. \ref{alpha_loci}). As for the other
response functions, we find two loci of strong extrema, minima in this case, $\alpha_P^{\rm
  smin}(P)$ and $\alpha_P^{\rm smin}(T)$, along isotherms
and isobars, respectively  showing a divergent behavior in the same region where we
find the strong maxima of $K_T$ and $C_P$. From this region two loci
of weaker minima depart. We find that the locus of weak minima along
isobars $\alpha_P^{\rm wmin}(T)$ bends toward the turning point of the
TMD. Although our calculations for $\alpha_P$ do not allow us to
observe the crossing with the TMD line,
based on the relation (see Methods)
\begin{equation}\label{rel4}
\left(\frac{\partial \alpha_P}{\partial T}\right)_{P,~\rm TMD}=-\frac{1}{V}\left(\frac{\partial P}{\partial T}\right)_{\rm TMD}\left(\frac{\partial^2V}{\partial P\partial T} \right)_{\rm TMD}
\end{equation} 
that holds at the TMD line, we can conclude that 
$\alpha_P^{\rm wmin}(T)$ should have zero $T$-derivative if it crosses the 
point where the TMD turns into the TminD line, because in this point
the TMD slope approaches zero.

The  locus of  weaker minima along isotherms $\alpha_P^{\rm wmin}(P)$, merges with the locus of maxima $\alpha_P^{\rm Max}(P)$ at the state point where the slope of both loci is $\partial P/\partial T\rightarrow\infty$ (not shown in Fig. \ref{alpha_loci}). According to the thermodynamic relation, discussed in Methods section,
\begin{equation}\label{rel3}
 \left(\dfrac{\partial\alpha_P}{\partial P}\right)_{T}=-\left(\dfrac{\partial K_T}{\partial T}\right)_P,
\end{equation}
we find that the locus of extrema in thermal expansivity along
isotherms coincides, within the error bars, with the locus of extrema
of isothermal compressibility along isobars (Fig. \ref{alpha_loci}c). 

All the loci of extrema of response functions that converge toward the same region $A$ in Fig. \ref{kt_loci}, \ref{cp_loci} and \ref{alpha_loci} increase in their absolute
values. Because  the increase of
response functions is related to the increase of fluctuations and this
is, in turn, related to the increase of correlation length $\xi$, to
estimate $\xi$ we
calculate the spatial correlation function  
\begin{equation}\label{eq_corr_func}
G(r)\equiv  \dfrac{1}{4N} \sum_{|\vec{r_i}-\vec{r_l}|=r} \left[\langle
  \sigma_{ij}(\vec{r_i}) \sigma_{lk}(\vec{r_l}) \rangle- \langle
  \sigma_{ij} \rangle^2\right] 
\end{equation} 
where $\vec{r_i}$ is the position of the molecule $i$,
$|\vec{r_i}-\vec{r_l}|=r$ the distance between molecule $i$ and
molecule $l$ and $ \langle \cdot \rangle$ the thermodynamic average.
The states of the water molecule, as well as the
density $\rho$, the energy $E$ and the entropy $S$ of the system, are
completely described by the bonding variables $\sigma_{ij}$. 
Therefore, the function $G(r)$ accounts for the fluctuations in $\rho$, $E$ and
$S$ and allows us to evaluate the correlation length because the order
parameter of the LLPT, as we discuss in the following, is related to a 
linear combination of $\rho$ and $E$. Note that, instead, the
density-density correlation function would give only an approximate
estimate of $\xi$.  

We observe an exponential decay of $G(r)\sim e^{-r/\xi}$ at high
temperatures in a broad range of pressures. Approaching the region
$A$, the correlation function can be written as $G(r)\sim
e^{-r/\xi}/r^{d-2+\eta}$ where $d$ is the dimension of the system and
$\eta$ a (critical) positive exponent. When
$\xi$ is of the order of the system size, the exponential factor
approaches a constant leaving the power-law as the dominant
contribution for the decay. 

\begin{figure}
\label{xi}
\includegraphics[scale=1,bb=70 310 612 720,clip=true]{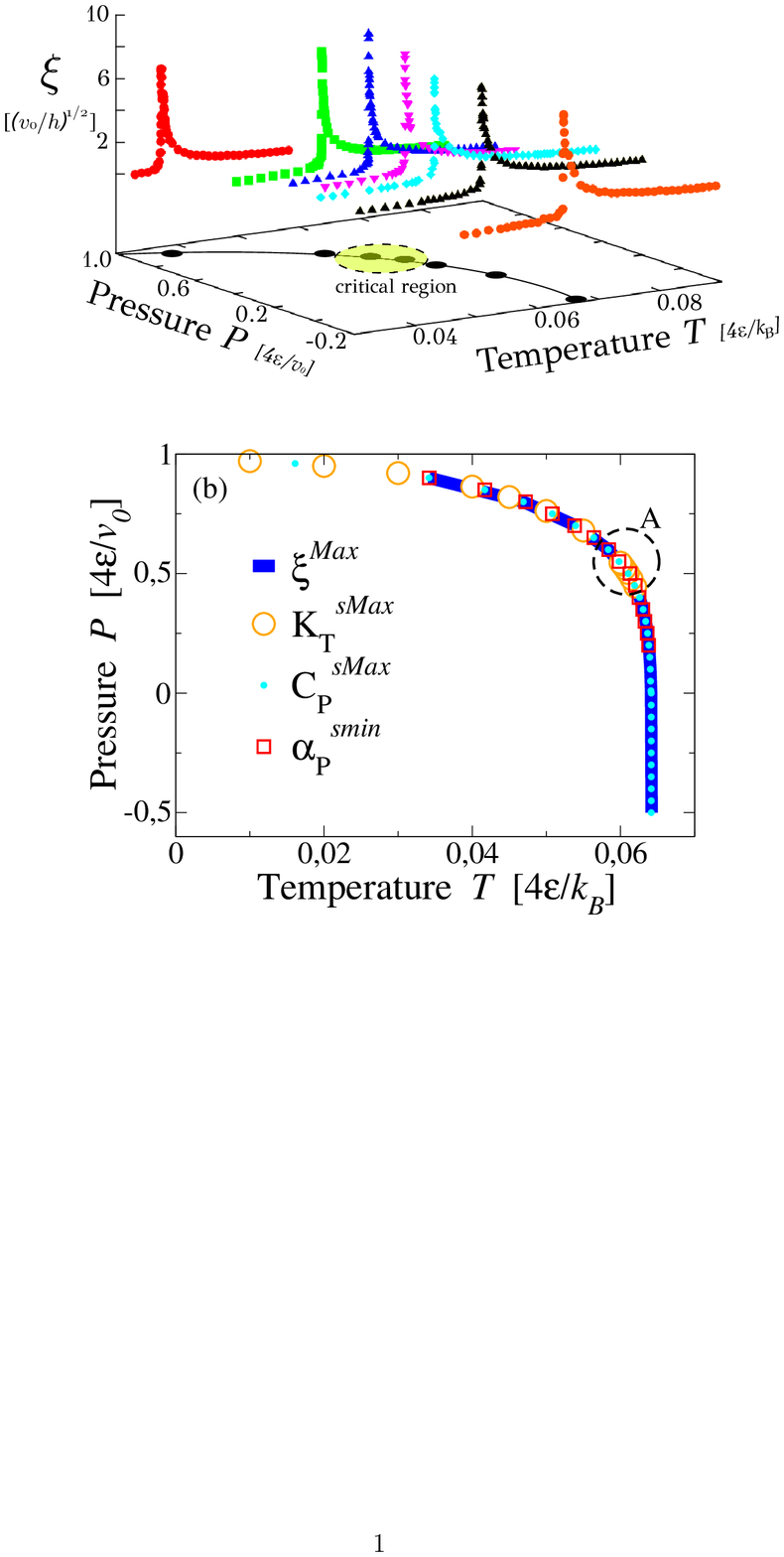}
\caption{(a) The correlation length $\xi$ along isobars for $N=10^4$ water
  molecules has maxima that increase for $P$ approaching the 
  critical region A.
 (b) The locus of $\xi$ maxima coincides with the loci of strong
extrema of $K_T$, $C_P$ and $\alpha_P$.
The Widom line is by definition the locus of 
  $\xi$ maxima at high $T$ departing from the LLCP, that we locate within the critical region A,
as discussed in the text.} 
\label{corr_length}
\end{figure}

At $P$ below the region $A$, we find that $\xi$ has a maximum, $\xi^{\rm Max}$, along isobars
and that  $\xi^{\rm Max}$ increases approaching 
$A$ (Fig. \ref{corr_length}). The $\xi^{\rm Max}$ locus 
coincides with the locus of strong extrema of 
$C_P$, $K_T$  and $\alpha_P$ 
(Fig. \ref{corr_length}b).  We observe that this common locus
converges to
$A$ and that all the extrema  increase approaching $A$. This behavior  is
consistent with the identification of $A$
with the critical region of the LLCP. 
Furthermore, we identify
the common locus with the Widom line 
that, by definition, 
is the $\xi^{\rm Max}$ locus departing from the LLCP in the one-phase region
\cite{xu, FS2007}. Our
calculations allow us to locate the Widom line at any $P$ down to the
liquid-to-gas spinodal.

At $P$ above the region $A$, we find the
continuation of the $\xi^{\rm Max}$ line, but with maxima that decrease for increasing $P$, as expected at the
LL spinodal that ends in the LLCP (Fig. \ref{corr_length}). 
Therefore, we identify
the high-$P$ part of the 
$\xi^{\rm Max}$ locus with the LL spinodal. 
Along this line the density, the energy and the entropy of the liquid
are discontinuous, as discussed in previous works
\cite{FMS2003,Stokely2010,KFS2008,dls,MSSSF2009,bb,bbb,Mazza}.

To better locate and characterize the LLCP in $A$ we need to define
the correct order parameter (o.p.) describing the LLPT. According to
mixed-field finite-size scaling theory  \cite{Wilding}, a
density-driven fluid--fluid phase transition is described by an o.p.
$M\equiv \rho^*+s u^*$, where $\rho*=\rho v_0$ represents the leading term
(number density), $u\equiv E/(\epsilon N)$ is the energy density (both
quantities are dimensionless)
and $s$ is the mixed-field parameter. Such linear combination is
necessary in order to get the rigth simmetry of the o.p.
distribution $Q_N(M)$ at the critical point where 
$Q_N(M)\propto \tilde{p}_d(x)$. Here is $x\equiv B(M-M_c)$, 
$B\equiv a_M^{-1}N^{\beta/d\nu}$, $\beta$ is the critical
exponent that governs $M$, $\nu$ is the critical exponent that governs
$\xi$, with $\nu$ and $\beta$
 defined by the universality class, $a_M$ is a
non-universal system-dependent parameter and $\tilde{p}_d$ is 
an universal function characteristic of the  Ising fixed--point in $d$ dimensions.
We adjust $B$ and $M_c$ so that $Q_N(M)$ has zero mean and unit variance. 

\begin{figure}
\includegraphics[scale=1,bb=70 260 612 720,clip=true]{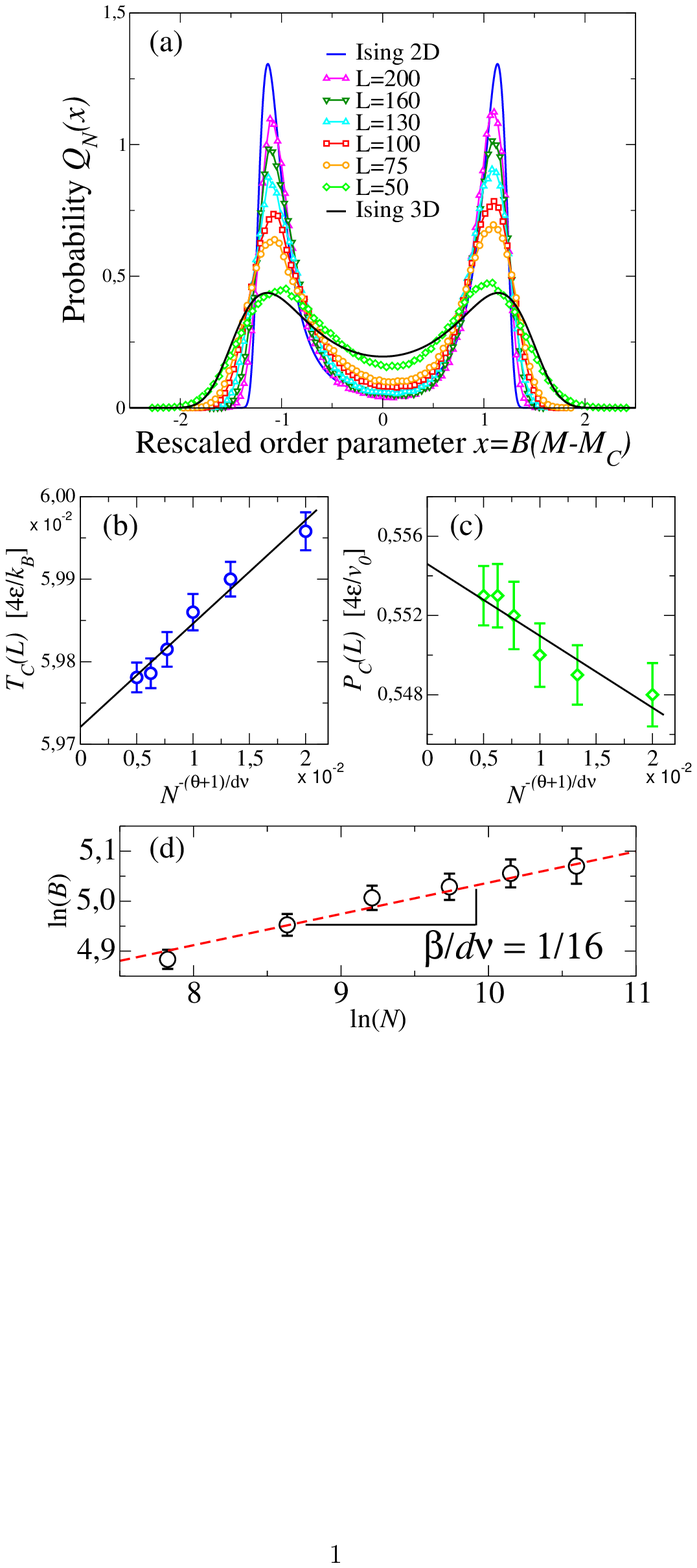}
\caption{(a) The size-dependent probability distribution $Q_N$ for the
  rescaled o.p. $x$, calculated for  $T_c(N)$, $P_c(N)$ and $B(N)$,
  has a symmetric shape that approaches continuously (from   $N=2500$,
  symbols at the top at $x=0$, to $N=40000$, symbols at the bottom)
  the limiting form for the 2D Ising universality class (full blue
  line) and differs from the 3D Ising universality class case (full black line). 
Error bars are smaller than the symbols size.
(b) The size-dependent LLCP temperature $T_c(N)$ and (c) pressure
$P_c(N)$  (symbols), resulting from our  best-fit of $Q_N$, 
extrapolate to  $T_c k_B/(4\epsilon)\simeq 0.0597$ and $P_c v_0/(4\epsilon)\simeq 0.555$, respectively, following the expected linear behaviors (lines).
(d) The normalization factor $B(N)$ (symbols) follows the power law function (dashed line) $\varpropto N^{\beta/d\nu}$. 
We use the $d=2$  Ising critical exponents: $\theta=2$ (correction to scaling), $\nu=1$ and $\beta=1/8$ (both defined in the text).} 
\label{critical_distributions}
\end{figure}

We combine, using the multiple histogram reweighting method \cite{hr1}
described in the Methods section,
a set of $3\times 10^4$ MC independent configurations
 for $\sim 300$ state points
with $0.040\leq T k_B/(4\epsilon)\leq 0.065$  and $0.40\leq Pv_0/(4\epsilon)\leq 0.75$. We verify,
by tuning $s$, $T$ and $P$,
that there is a point within the region $A$ where the calculated
$Q_N(x)$ has a symmetric shape with respect to $x=0$
(Fig.~\ref{critical_distributions}). We find $s=0.25\pm 0.03$ for our
range of $N$. The resulting critical parameters $T_c(N)$, $P_c(N)$ and
the normalization factor $B(N)$ follow the expected
finite-size behaviors with 2D Ising critical exponents
\cite{Wilding}. From the finite-size analysis  we extract the
asymptotic values $T_c k_B/(4\epsilon)=0.0597\pm 0.0001$  and $P_c v_0/(4\epsilon)=0.555\pm 0.002$.  

\begin{figure}
\centering
 \includegraphics[scale=0.55,bb=2 0 754 613,clip=true]{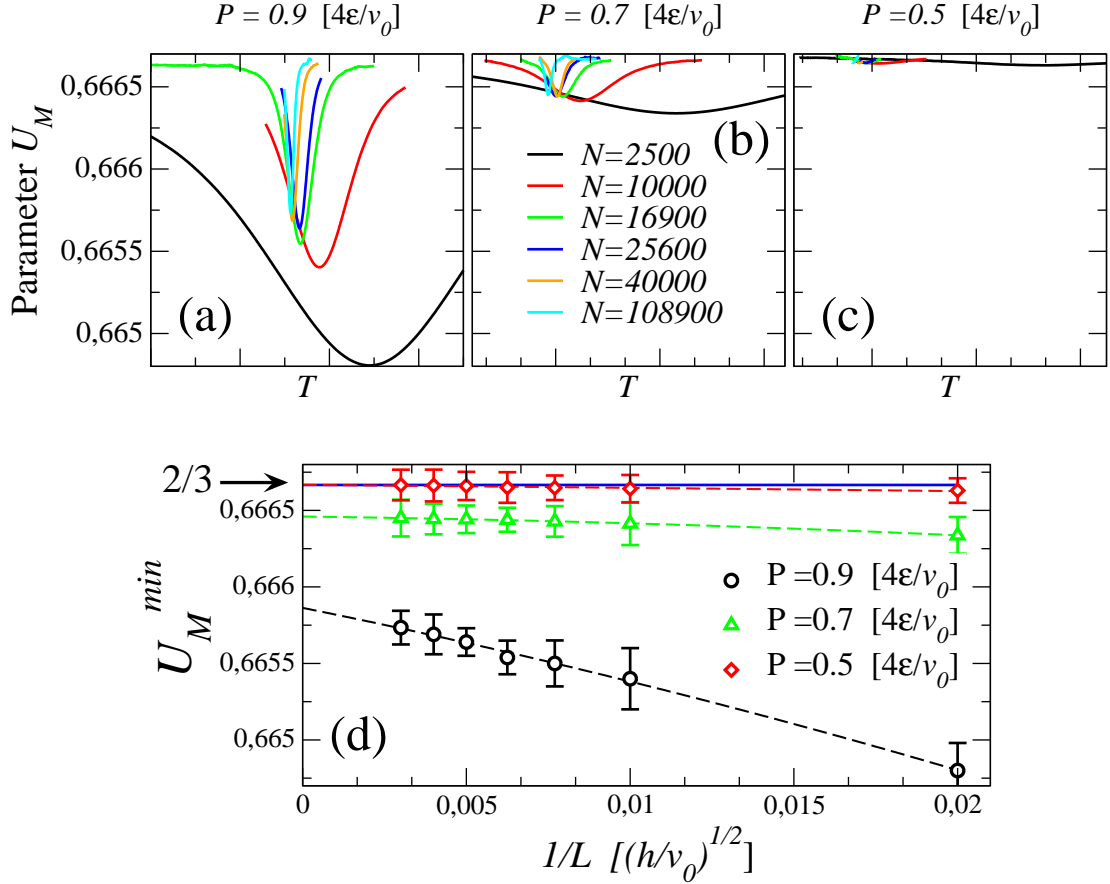}
\caption{Challa-Landau-Binder parameter $U_M$ (defined in the text) of
the o.p. $M$ for different system sizes, calculated for three
pressures: (a) $P v_0/(4\epsilon)=0.9$, (b) $P v_0/(4\epsilon)=0.7$, and (c) $P v_0/(4\epsilon)=0.5$ slightly below
$P_c v_0/(4\epsilon)\simeq 0.555$ . The curves are calculated with the histogram reweighting method. (d) Scaling of the minima of $U_M$ for different $P$. The arrow points to value  $2/3$ corresponding to the absence of a first-order phase transition in the thermodynamic limit. Error bars are calculated propagating the statistical error from histogram reweighting method.} 
\label{binder}
\end{figure}

The presence of a first order phase transition ending in a critical
point, associated to the o.p. $M$, is confirmed by the finite size
analysis of the Challa-Landau-Binder parameter  \cite{cumulant} of $M$ 
\begin{equation}
 U_M\equiv 1-\dfrac{\langle M^4\rangle_N}{3\langle M^2\rangle^2_N}\qquad
\end{equation} 
where the symbol $\langle \cdotp\rangle_N$ refers to the thermodynamic
average for a system with $N$ water molecules. $U_M$ quantifies the
bimodality in $Q_N(M)$. The isobaric value of $U_M$ shows a minimum at
the temperature where $Q_N(M)$ mostly deviates with respect to a
symmetric distribution (Fig. \ref{binder}). Minimum of $U_M$
converges to $2/3$ in the thermodynamic limit away from a first order
phase transition, while it approaches to a value $<2/3$ 
where the bimodality of $Q_N(M)$ indicates the
presence of phase coexistence.  

\begin{figure}
 \includegraphics[scale=0.5,bb=14 3 750 562,clip=true]{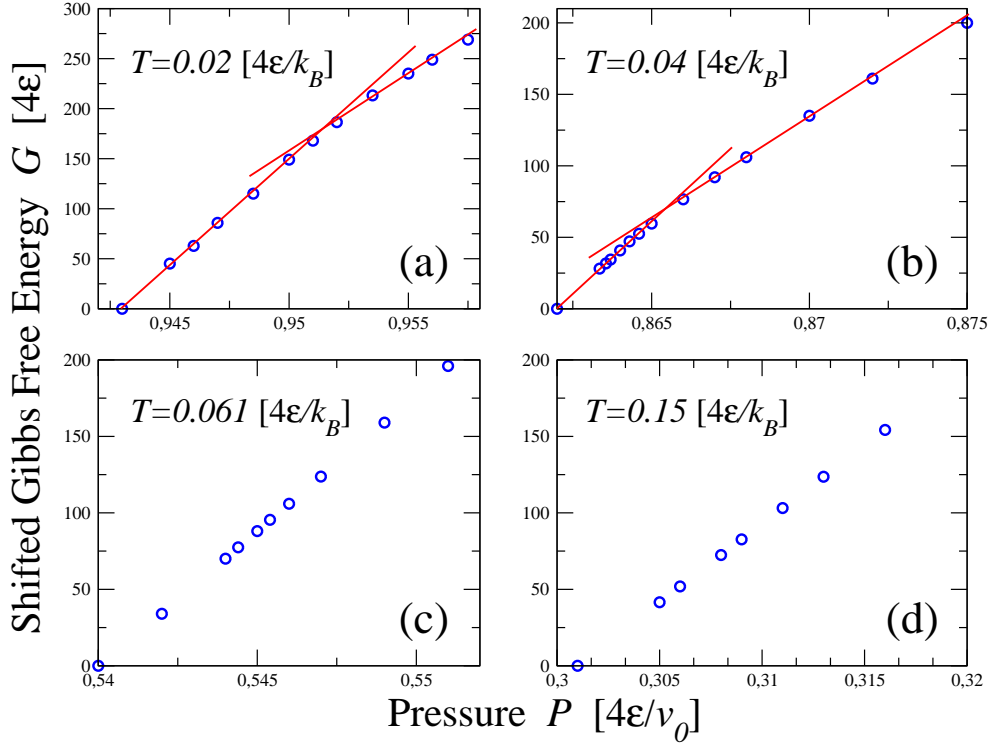}
\caption{Gibbs free energy $G$ along isotherms, as
    function of   $P$. Points are shifted in such a way that $G=0$ at the lowest
    $P$. Lines are   guides for the eyes. 
(a) For $T=0.02 (4\epsilon/k_B)<T_c$ there is a discontinuity in the $P$-derivative of $G$ at $P\simeq 0.952 (4\epsilon/v_0) >P_c$ as expected at the LLPT, consistent with the behavior of the response functions at this state point (e.g., in Fig.~3b, 4b).  (b) For $T=0.04 (4\epsilon/k_B)<T_c$ we observe the discontinuity in the $P$-derivative at $P\simeq 0.865 (4\epsilon/v_0) >P_c$, again consistent with the LLPT. The LDL has a lower chemical potential
    ($\mu \equiv G/N$) than the HDL, $\mu_{LDL}<\mu_{HDL}$, due to the
    HB energy gain in the LDL. 
For $T=0.061 (4\epsilon/k_B)$ (c) and for  $T=0.15 (4\epsilon/k_B)$ (d), both larger than $T_c$,  we instead do not observe any discontinuity in the $P$-derivative of $G$ by crossing the locus of $C_P(P)^{\rm wMax}$ and the locus of $K_T(T)^{\rm wMax}$, respectively, as expected in the one-phase region.
} 
\label{gibbs_energy}
\end{figure}

These results are consistent with the behavior of the Gibbs free energy
$G$ calculated with the histogram reweighting method
(Fig. \ref{gibbs_energy}).   In particular, we calculate $G$ along isotherms, for $P$ crossing the LLPT and the loci of weak maxima in
$K_T(T)$ and $C_P(P)$. We find that the behavior of $G$ for $T < T_c$
is consistent with the occurrence of a discontinuity in volume $V
= \partial G/\partial P$,  in the thermodynamic limit, with a decrease of $V$ corresponding to the transition from LDL to HDL for increasing $P$. Crossing the loci $K_T(T)^{\rm wMax}$ and $C_P(P)^{\rm wMax}$ the volume decreases with pressure without any discontinuity as expected in the one-phase region.

\begin{figure}
\includegraphics[scale=1,bb=70 395 612 720,clip=true]{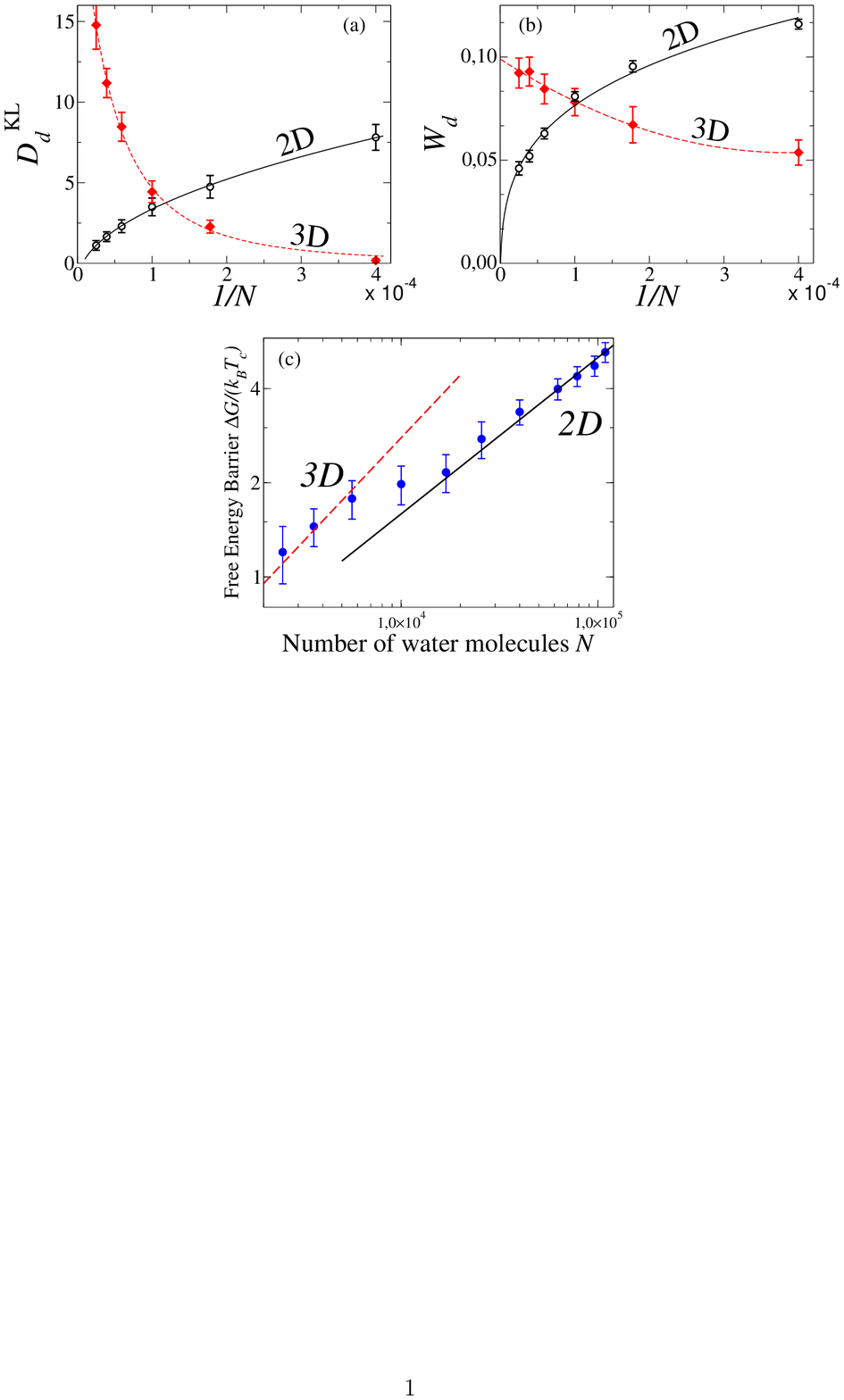}
\caption{(a) Kullback-Leibler divergence $D^{\rm KL}_d(N)$ and (b) Liu
  et al. deviations $W_d$  of the
  calculated $\tilde{p}(N)$ from the Ising universal function
  $\tilde{p}_d$ in $d=2$ (open symbols) and $d=3$
  (closed symbols), as a function of $1/N$, with $N$ 
water molecules, at constant
  $\rho\simeq \rho_c$. 
In both panels lines are power-law fits and 
we observe a crossover between 2D and 3D behavior at $N\simeq 10^4$.
(c) The free-energy cost to form an interface between the two liquids
coexisting at the LLCP scales as  $\Delta G \propto N^{\frac{d-1}{d}}$
with $d=3$ for $N<10^4$ and  $d=2$ for $N>10^4$. }   
\label{W}
\end{figure}

The distribution $Q_N(N)$ adjust well to the data only for large
$N$. We, therefore, perform a more systematic analysis. For each $N$,
we quantify the deviation of the calculated $\tilde{p}(N)$ from the
expected $\tilde{p}_{2}$ for the 2D Ising. Furthermore, due to the
behavior of data for small $N$  (Fig.~\ref{critical_distributions}a),
we calculate the deviation 
from the 3D Ising $\tilde{p}_{3}$ \cite{Wilding}.
We estimate the Kullback-Leibler divergence \cite{kullback-leibler},
\begin{equation}\label{scarto_kl}
D^{\rm KL}_d(N) \equiv 
\sum_{i=1}^{n} \ln \left(
  \dfrac{\tilde{p}_{d,i}}{\tilde{p}_i(N)}\right) \tilde{p}_{d,i}
\end{equation} 
of the probability distribution $\tilde{p}_i(N)$ of $x_i$ from the
theoretical value $\tilde{p}_{d,i}$ of $x_i$ ($i=1, \dots, n$) in $d$
dimensions (Fig.~\ref{W}a),
and the Liu et al. deviation \cite{Liu2010},
\begin{equation}\label{scarto_2}
W_d(N) \equiv
\dfrac{1}{n}\dfrac{\sum_{i=1}^{n}\sqrt{\tilde{p}_i(N)}|\tilde{p}_i(N)-\tilde{p}_{d,i}|}{\tilde{p}_{d,{\rm      peak}}-\tilde{p}_{d,x=0}}
\end{equation} 
with $\tilde{p}_{d,{\rm peak}}-\tilde{p}_{d,x=0}$ 
difference between the distribution peak and its
value at $x=0$ (Fig.~\ref{W}b).

We confirm $s\simeq 0.25$ for $\tilde{p}_2$ and find $s=0.10\pm 0.02$ for $\tilde{p}_3$ for our range of $N$. For both $D^{\rm KL}_d$ and $W_d$, with $d=2$ and $d=3$, we find minima at $T_c  k_B/(4\epsilon)\simeq 0.06$ and $P_c  v_0/(4\epsilon)\simeq 0.55$
that become stronger for increasing $N$.
We find that $D^{\rm KL}_2$ and $W_2$ decrease with increasing $N$, vanishing  for $N\rightarrow \infty$ 
(Fig.~\ref{W}). Therefore, for an infinite monolayer between
hydrophobic walls separated by $h\approx 0.5$~nm, the system has a
LLCP that belongs to the 2D Ising universality class, as expected from
our representation of the system as the 2D projection of the monolayer.  

However, by increasing the confinement,  i.e. reducing $N$ and $L$ at
constant $\rho$, $D^{\rm KL}_2$ and $W_2$ become larger than $D^{\rm
  KL}_3$ and $W_3$, respectively. Therefore, the calculated
$\tilde{p}(N)$ deviates from the 3D probability distribution less than
from the 2D probability distribution. For $N=2500$ we find that both
$D^{\rm KL}_3$ and $W_3$ have values approximately equal to those for
$D^{\rm KL}_2$ and $W_2$ calculated for a system ten times larger. In
particular we find $D^{\rm KL}_3\simeq 0$ for $N=2500$. Hence, by
increasing the confinement of the monolayer at constant $\rho$, the
LLCP has a behavior that 
approximates better the bulk~\cite{LLCPnew, llcp1, llcp2, llcp3, 
  llcp4,llcp4_bis, llcp5}, with a crossover between 2D and 3D-behavior occurring
at $N\simeq 10^4$. 

This dimensional crossover is confirmed by the finite-size analysis of
the Gibbs free energy cost $\Delta G/(k_BT_c)$ to form an interface
between the two liquids in the vicinity of the LLCP, calculated as
$\Delta G(N) \equiv -k_BT_c(N)[\ln \mathscr{P}_N^{\rm min}(\mathscr{H},V) -\ln \mathscr{P}_N^{\rm Max}(\mathscr{H},V)]$, 
where $\mathscr{P}_N^{\rm min}$ and $\mathscr{P}_N^{\rm Max}$
are the minimum and maximum values of the probability distribution 
$\mathscr{P}_N(\mathscr{H},V)$ of
configurations of $N$ water molecules with energy $\mathscr{H}$ and
volume $V$ at the LLCP.
This quantity is expected to scale as $\Delta G \propto 
N^{\frac{d-1}{d}}$. We find that our  data can be fitted as $N^{\frac{2}{3}}$ for small sizes and as $N^{\frac{1}{2}}$ for large sizes with a crossover around $N=10^4$ (Fig.~\ref{W}c). Considering the value of the estimated $\rho_c$ in real units ($\simeq
1$g/cm$^3$) \cite{dls}, the corresponding crossover wall-size is $L\simeq 25$~nm.

\medskip
\bigskip

\noindent{\bf Discussion}

Our rationale for this dimensional crossover at fixed $h$ is that,
when $L/h$ decreases toward 1, the characteristic way the critical fluctuations
spread over the system, i.e. the universality class of the LLCP,
resembles closely the bulk because the asymmetry among the three
spatial dimensions is reduced. A similar result was found recently by
Liu et al. for the gas-liquid critical point of a Lennard-Jones (LJ)
system confined between walls by fixing $L$ and varying $h$
\cite{Liu2010}. However, in the case considered by Liu et al. the
crossover was expected because  the number of layers of particles was
increased from one to several, making the system more similar to the
isotropic 3D case. Here, instead, we consider always one single layer,
changing the proportion $L/h$ by varying $L$. Therefore, it could be
expected that the system belongs 
to the 2D universality class for any $L$.  

Furthermore, the extrapolation of the results for the
  LJ liquid to our case of a monolayer with $h/r_0\simeq 1.7$, where
  $r_0$ is the water van der Waals diameter, would predict a
  dimensional crossover at $L/h\simeq 5$ \cite{Liu2010}.  Here,
  instead, we find the crossover at $L/h\simeq 50$, 
  i.e. one order of magnitude larger than the LJ case. We ascribe this
  enhancement of the crossover to (i) the presence of a cooperative HB
  network and (ii) the low coordination number that water has in both
  the monolayer and the bulk. These are
the main differences between water and a LJ
  fluid. The cooperativity intensifies drastically the spreading of
  the critical fluctuations along a network, contributing to the 
effective
  dimensionality increase of the confined monolayer. 
Moreover, the HB network has in 3D a 
 coordination number ($z=4$) as low as in 2D, making the 
first coordination shell similar in both dimensions. 

Our  findings are consistent with recent atomistic simulations of
water nanoconfined between surfaces.
\cite{zhang, Ballenegger, Bonthuis}.
Zhang et al. found that water dipolar fluctuations are enhanced
in the direction parallel to the confining surfaces (hydrophobic graphene
sheets) within a distance
of 0.5~nm \cite{zhang}. Ballenegger and Hansen 
found similar results for confined
polar fluids, including water, within $\approx  0.5$~nm distance from
the hydrophobic surface \cite{Ballenegger}. Bonthuis et al. extended
these results to both hydrophilic and hydrophobic confining surfaces.
All these findings are 
 consistent with our result showing the
enhancement of the fluctuations of the o.p. in the direction parallel
to the confining walls separated by $h\approx 0.5$~nm. 
Furthermore, 
Zhang et al. observed that the effect does not depend on the
details of the water-surface interaction but stems from the very
presence of interfaces \cite{zhang}. This is confirmed by our study, where the
water-interface interaction is purely due to excluded volume.
Following the authors of Ref. \cite{zhang}, this observation allows us
to relate our finding for rigid surfaces to experimental results for 
water hydrating membranes \cite{Tielrooij}, reporting new types of
water dynamics in thin interfacial layers,  and water nanoconfined in
different types of reverse micelles \cite{Moilanen}, showing that the
water dynamics is governed by the presence of the interface 
rather than the details (e.g.,  the presence charged
groups) of the interface.

In conclusion, we analyze the low-$T$ phase diagram of a water
monolayer confined between hydrophobic parallel walls of size $L$
separated by $h\approx 0.5$~nm. 
We study water fluctuations associated to the thermodynamic
response functions and their relations to the loci of TMD, TminD. For
each response function  we find two loci of extrema, one  
stronger at lower-$T$ and one weaker and broader at higher-$T$. 
These loci converge toward a critical region where the fluctuations
diverge in the thermodynamic limit, defining the LLCP. We calculate
the Widom line departing from the LLCP based on its definition as the
locus of maxima of $\xi$ and show that it coincides with the locus of
strong maxima of the response functions. 
We find that the LLCP  belongs to the 2D Ising universality class for
$L\rightarrow \infty$, with strong finite-size effects for small $L$.
Surprisingly, the finite-size effects induce the LLCP universality
class to converge toward the bulk case 
(3D Ising universality class)
already for a system with a very
pronounced plane asymmetry, i.e. a water monolayer of height
$h\approx 0.5$~nm and $L/h \approx 50$. For normal liquid, instead, 
this is expected only for  much smaller relative values of $L$
($L/h\leq 5$).  
We rationalize this result  as a
consequence of two properties of the HB 
network: (i) its high cooperativity, that enhances the fluctuations, and (ii) its low coordination
number, that makes the first coordination shell for the 
monolayer and the bulk similar.

\medskip
\bigskip

\noindent{\bf Methods}

{\it The Model.}  We consider a monolayer formed by $N$ water molecules confined in a
volume $V\equiv hL^2$ between two hydrophobic flat surfaces separated
by a distance $h$,  with  $V/N\geq v_0\simeq 42$ \AA{}$^3$, where $v_0$ is the water 
excluded volume. 
Each water molecule has four next-neighbours \cite{shr}. We partition the volume into $N$ equivalent
cells of height $h\backsimeq 0.5 $nm and square section with size
$r\equiv \sqrt{L^2/N}$, equal to the average distance between water
molecules.
By coarse-graining the molecules distance from the
surfaces, we reduce our monolayer representation to a 2D system. We
use periodic boundary conditions parallel to the walls to reduce
finite-size effects. We simulate constant $N$, $P$, $T$, allowing
$V(T,P)$ to change, with each cell $i=1, \dots, N$ 
having number density $\rho_{i}\equiv
\rho(T,P)\equiv N/V\leq \rho_0\equiv 1/v_0$ corresponding to a mass density $\backsimeq 1$  g/cm$^3$. To each cell we associate
a variable $n_{i}=0$ ($n_{i}=1$) depending if the cell $i$  has
$\rho_{i}/\rho_0\leq 0.5$ ($\rho_{i}/\rho_0> 0.5$). Hence, $n_{i}$ is
a discretized density field replacing the water translational degrees
of freedom. 
The water-water
interaction is given by  
\begin{equation}\label{hamiltoniana}
 \mathscr{H}\equiv \sum_{ij} U(r_{ij}) -JN_{\rm HB} - J_{\sigma}N_{\rm
   coop}.
\end{equation} 
The first term, summed over all the water molecules $i$ and $j$ at O--O distance $r_{ij}$, has $U(r)\equiv \infty$
for $r<r_0\equiv \sqrt{v_0/h} = 2.9 $ \AA{} (water van der Waals diameter), $U(r)\equiv 4\epsilon
[(r_0/r)^{12}-(r_0/r)^6]$ for $r\geq r_0$ with  $\epsilon \equiv 5.8$ kJ/mol, and $U(r)\equiv 0$ for
$r>r_c \equiv 25 r_0$ (cutoff).

The second term represents the directional (covalent) component of the
hydrogen bond (HB), with 
$J/4\epsilon\equiv 0.5$, $N_{\rm HB}\equiv\sum_{\langle ij \rangle}n_i
n_j\delta_{\sigma_{ij},\sigma_{ji}}$ number of HBs, with the sum over
n.n., where $\sigma_{ij}=1, \dots, q$ is the bonding index of molecule $i$ to the
n.n. molecule $j$, with $\delta_{ab}=1$ if $a=b$, 0 otherwise. Each water
molecule can form up to four HBs. 
We adopt a geometrical definition of the HB, based on the
${\widehat{\rm OOH}}$ angle and the OH---O distance.
A HB breaks if  ${\widehat{\rm OOH}}> 30^o$. Hence, only $1/6$ of 
the entire range of values $[0,360^\circ]$ for  the
${\widehat{\rm OOH}}$ angle is associated to a bonded state.
Therefore, we choose $q=6$ to account correctly for the entropy
variation due to the HB 
formation and breaking.
Moreover, a  HB breaks when the OH---O distance $>r_{\rm max}-r_{\rm
  OH}=3.14$ \AA{}, where $r_{\rm OH}=0.96$ \AA{} and $r_{\rm max} =4.1
$ \AA{}. The value of $r_{\rm max}$ is a consequence of our choice
$n_i=0$ for $\rho_i/\rho_0\leq 0.5$, i.e. $r_i^2/2\geq r_0^2$,
implying that $n_i n_j =0$ when $r_{ij}\geq r_0\sqrt{2}=4.10$ \AA{} $\equiv
r_{\rm max}$.

The third term of the Eq.(\ref{hamiltoniana}) accounts for  the HB
cooperativity due to the quantum many-body interaction
\cite{Hernandez}, with $J_\sigma/4\epsilon\equiv 0.05$ and
$N_{\rm coop}\equiv \sum_i
n_i\sum_{(l,k)_i}\delta_{\sigma_{ik},\sigma_{il}}$,  
where $(l,k)_i$ indicates each of the six different pairs of the four
indices $\sigma_{ij}$ of a molecule $i$. 
The value $J_\sigma\ll J$ is chosen in such a way to guarantee an
asymmetry between the two components of the HB interaction.
To the cooperative term is due the   
O--O--O correlation that locally leads the molecules toward an ordered
configuration. 
In bulk water this term would lead to a tetrahedral structure at low
$P$  up to the second shell,  as observed in the experiments
\cite{SoperRicci2000}. An increase of $T$ 
or $P$ partially disrupts the HB network and induces a more compact
local structure, with smaller average volume per molecule. Therefore,
for each HB we include an enthalpy increase $Pv_{\rm HB}$,  where
$v_{\rm HB}/v_0=0.5$ is the average volume increase between
high-$\rho$ ices VI and VIII and low-$\rho$ (tetrahedral) ice
Ih. Hence, the total volume is $V\equiv V_0+N_{\rm HB}v_{\rm HB}$,
where $V_0\geq Nv_0$ is a stochastic continuous 
variable changing with Monte Carlo (MC) acceptance rule \cite{MSSSF2009}. Because the HBs do not affect the n.n. distance \cite{SoperRicci2000},
we ignore their negligible effect on the $U(r)$ term. 
 Finally, we model the water-wall interaction by excluded volume. 

{\it The Observables}. The LLCP is identified by the mixed-field order parameter $M$ and not by the magnetization of the Potts variables $\sigma_{i,j}$ as in normal Potts model. $M$ is related to the configuration of the system by the relation
\begin{equation}
M\equiv \dfrac{N}{ Nv + \sum_{\langle ij \rangle}n_i n_j\delta_{\sigma_{ij},\sigma_{ji}}} + s\left( U(r) -J\sum_{\langle ij \rangle}n_i n_j\delta_{\sigma_{ij},\sigma_{ji}} - J_\sigma \sum_i n_i\sum_{(l,k)_i}\delta_{\sigma_{ik},\sigma_{il}} \right)
\end{equation}
where $v\equiv V_0/N$  and $s$ is the mixed-field parameter. $M$ is therefore a linear combination of the density and energy.

Thermodynamic response functions are calculated from 
\begin{equation}\label{kt}
K_T\equiv- \frac{1}{\langle V \rangle}\left(\frac{\partial \langle V \rangle}{\partial P}\right)_T = \frac{\langle\Delta V^2\rangle}{k_BTV}
\end{equation}
and 
\begin{equation}\label{cp}
 C_P\equiv \left(\frac{\partial \langle H \rangle}{\partial T}\right)_P =\frac{\langle\Delta H^2\rangle}{k_BT}
\end{equation}
as long as the volume and energy distributions are not clearly bimodal, i.e. excluding the values of $T$ and $P$ where the phase coexistence is observed, based on the definition of $M$. Here $\Delta O \equiv  O -  \langle O\rangle$, for $O=V, H$ and, $H$ is the enthalpy of the system. 

{\it The Monte Carlo Method}.
The system is equilibrated via Monte Carlo simulation with Wolff algorithm \cite{MSSSF2009}, following an annealing procedure: starting with random initial condition at high $T$, the temperature is slowly decreased and the system is re-equilibrated and sampled with $10^4\div 10^5$ independent configurations for each state point.                                                                
The thermodynamic equilibrium is checked probing that the fluctuation-dissipation relations, Eq. (\ref{kt}) and (\ref{cp}), hold within the error bar.

{\it The Histogram Reweighting Method}. 
The probability $Q_N(M)$ is calculated in a continuous range of $T$ and $P$ across the $\xi^{\rm Max}$ line.
We consider an initial set of $m\in [10:20]$ independent simulations within a temperature range $\Delta T k_B/(4\epsilon)\sim10^{-4}$ and a pressure range $\Delta P v_0/(4\epsilon)\sim 10^{-3}$. For each simulation $i=1, ..., m$ we calculate the histograms $h_i(u,\rho)$ in the energy density--density plane. The histograms $h_i(u,\rho)$ provide an estimate of the equilibrium probability distribution for $u$ and $\rho$; this estimate becomes correct in the thermodynamic limit.
For the $NPT$ ensemble, the new histogram $h(u,\rho,P',\beta')$ for new values of $\beta'=1/k_BT'$ and $P'$ close the simulated ones, is given by the relation \cite{hr1}
\begin{equation}\label{hr1}
h(u,\rho,P',\beta')=\dfrac{\sum_{i=1}^m h_i(u,\rho) e^{-\beta'(u+P' /\rho)N}}{\sum_{i=1}^m N_i  e^{-\beta_i(u+P_i/\rho)N - C_i}}
\end{equation} 
where $N_i$ is the number of independent configurations of the run $i$.
The constants $C_i$, related to the Gibbs free energy value at $T_i$ and $P_i$, are self-consistently calculated from the equation \cite{hr1}
\begin{equation}\label{hr2}
e^{C_i}=\sum_u\sum_\rho h(u,\rho,P_i,\beta_i)\simeq Z(P_i,\beta_i)\qquad \Rightarrow \qquad C_i=-G(P_i,\beta_i)/k_BT\text{ }.
\end{equation}
We choose as initial set of parameters $C_i=0$. The parameters $C_i$ are recursively calculated by means of Eq. (\ref{hr1}) and (\ref{hr2}) until the difference between the values at iteration $k$ and $k+1$ is less then the desired numerical resolution ($10^{-3}$ in our calculations).
Once the new histogram is calculated, $Q_N(M)$ at $T_i$ and $P_i$ is calculated integrating $h(u,\rho,P_i,\beta_i)$ along a direction perpendicular to the line $\rho+su$.

{\it Thermodynamic relations}. 
We report here the calculations for the thermodynamic relations in Eq. (\ref{rel1}), (\ref{rel2}), (\ref{rel4}) and (\ref{rel3}) \cite{sf}.
To verify the relation (\ref{rel3}) we calculate the derivative of $K_T$ along isobars
\begin{equation}
\begin{array}{ll}
\left(\dfrac{\partial K_T}{\partial T}\right)_P = \dfrac{\partial}{\partial T}\left(-\dfrac{1}{V} \left(\dfrac{\partial V}{\partial P}\right)_T \right)_P
=\dfrac{1}{V^2}\left(\dfrac{\partial V}{\partial T}\right)_P\left(\dfrac{\partial V}{\partial P}\right)_T - \dfrac{1}{V}\dfrac{\partial^2 V}{\partial P\partial T} = -\alpha_P K_T - \dfrac{1}{V}\dfrac{\partial^2 V}{\partial P\partial T} 
\end{array} 
\label{kt/t}
\end{equation}
and the derivative of $\alpha_P$ along isotherms
\begin{equation}
\begin{array}{ll}
\left(\dfrac{\partial \alpha_P}{\partial P}\right)_T & = \dfrac{\partial}{\partial P}\left(\dfrac{1}{V} \left( \dfrac{\partial V}{\partial T}\right)_P\right)_T = -\dfrac{1}{V^2}\left(\dfrac{\partial V}{\partial P}\right)_T\left(\dfrac{\partial V}{\partial T}\right)_P + \dfrac{1}{V}\dfrac{\partial^2 V}{\partial T\partial P}= \\\\ &=\alpha_P K_T+ \dfrac{1}{V}\dfrac{\partial^2 V}{\partial P\partial T}  = -  \left(\dfrac{\partial K_T}{\partial T}\right)_P 
\end{array} 
\end{equation}

Following \cite{sf, rebelo} the line of extrema in density (TMD and TminD lines)
is characterized by 
$\alpha_P=0$, hence, $d\alpha_P=0$ along the TMD line. 
Therefore,
\begin{equation}\label{tmd}
\begin{array}{ll}
0=d\alpha_P\equiv \left(\dfrac{\partial \alpha_P}{\partial T}\right)_{P,~\rm ED}dT + \left(\dfrac{\partial \alpha_P}{\partial P}\right)_{T,~\rm ED}dP  
=\left( \dfrac{1}{V} \dfrac{\partial^2 V}{\partial T^2}\right)_{P,~\rm ED}dT +  \left( \dfrac{1}{V}\dfrac{\partial^2 V}{\partial P\partial T} \right)_{\rm ED}dP
\end{array}
\end{equation}
where the index ``ED'' denotes that the derivatives are taken along the locus of extrema in density.
So, the slope $\partial P/\partial T $ of TMD is given by
\begin{equation}\label{tmd1}
 \left(\frac{\partial P}{\partial T}\right)_{\rm TMD} = - \dfrac{\left( \dfrac{\partial^2 V}{\partial T^2} \right)_{P,~\rm TMD}}
{ \left( \dfrac{\partial^2 V}{\partial P\partial T } \right )_{\rm TMD}}
\end{equation} 
from which, using Eq. (\ref{kt/t}) with $\alpha_P=0$, we get Eq. (\ref{rel1}). The Eq. (\ref{tmd1}) holds  as long as both $(\partial \alpha_P/\partial P)_T$ and $(\partial \alpha_P/\partial T)_P$ do not vanish contemporary, as it occurs along the Widom line, where the loci of strong minima of $\alpha_P$ overlap. 
For this reason the intersection between the Widom line and TminD line does not imply any change in the slope $(\partial P/\partial T)_{\rm TminD}$. 

To calculate Eq. (\ref{rel2}) we start from $C_P$ and $\alpha_P$ written in terms of Gibbs free energy
\begin{equation}
 \frac{C_P}{T}=-\frac{\partial^2 G}{\partial T^2}\qquad , \qquad V\alpha_P = \frac{\partial^2 G}{\partial P \partial T}
\end{equation} 
from which results
\begin{equation}\label{eq}
 \begin{array}{ll}
\dfrac{\partial}{\partial P}\left(\dfrac{C_P}{T} \right)_T  = \dfrac{1}{T}\left(\dfrac{\partial C_P}{\partial P}\right)_T = - \left[\dfrac{\partial}{\partial T} \left( V\alpha_P\right)\right]_P 
 = -\left(\dfrac{\partial V}{\partial T}\right)_P\alpha_P-V\left(\dfrac{\partial \alpha_P}{\partial T}\right)_P=
-\left(\dfrac{\partial^2 V}{\partial T^2}\right)_P\qquad ,
 \end{array}
\end{equation} 
\begin{equation}
 \left(\dfrac{\partial C_P}{\partial P}\right)_T = -T\left( \dfrac{\partial^2 V}{\partial T^2}\right)_P\qquad .
\end{equation}

Substituting in Eq. (\ref{tmd1}) we get the Eq. (\ref{rel2}) at the TMD. Moreover, because of $\alpha_P=0$ at the TMD line, from the last equivalence of Eq. (\ref{eq}) we get
\begin{equation}
\left(\dfrac{\partial \alpha_P}{\partial T}\right)_{P,~\rm TMD} = \dfrac{1}{V} \left(\dfrac{\partial^2 V}{\partial T^2}\right)_{P,~\rm TMD}
\end{equation}
from which, using Eq. (\ref{tmd1}), we get the Eq. (\ref{rel4}).

\bigskip

\medskip
\bigskip
\noindent{\bf ACKNOWLEDGEMENTS}

\noindent  We acknowledge the support of 
Spanish MEC grant FIS2012-31025  and the EU FP7 grant NMP4-SL-2011-266737.
\newpage

\noindent{\bf AUTHOR CONTRIBUTION STATEMENT}

\noindent  
V. B. and G. F. designed the research. V. B. made the simulations. G.
F. supervised the work. Both authors analyzed the data, prepared the
figures, wrote the text and reviewed the manuscript.

\newpage


\medskip
\bigskip
\noindent{\bf ADDITIONAL INFORMATION}

The authors declare no competing financial interests.

\newpage


\end{document}